\documentclass[twocolumn,aps,prl,showpacs,preprintnumbers,amsmath,amssymb]{revtex4}
\usepackage{epsfig}
\usepackage{amssymb,amsfonts}

\begin{document}
\def\mydag{^{\vphantom{\dagger}}}
\title{Transport and Quantum Walk of Nonclassical Light
in Coupled Waveguides}

\author{ Amit Rai, G. S. Agarwal and J. H. H. Perk} \affiliation{Department of Physics,
Oklahoma State University, Stillwater, Oklahoma 74078, USA}

\date{\today}

\begin{abstract}
We study the transport and quantum walk of nonclassical light in an
array of coupled wave\-guides which have novel properties like very
low decoherence and thus making them ideal for storage of quantum
information. We show how squeezing gets turned over from one
waveguide to another. We further show how input nonclassical light
can generate entanglement among different wave\-guides. Our results
involve both first quantization due to array structure  and second
quantization due to the quantum nature of fields and can also be
used to discuss the Talbot effect in the quantum regime.
\end{abstract}


\maketitle Optical elements like beam splitters are known to behave
quite differently when it comes to single photons. A single photon
according to Dirac can either be transmitted or reflected
\cite{dirac}. It can not be found simultaneously in both
transmission or reflection. The behavior of a photon pair on a beam
splitter is even more remarkable as shown by Hong, Ou and Mandel
\cite{hong}. For a 50-50 beam splitter both photons are found either
in transmission or in reflection. This leads to entanglement of
photons at the two output  ports. The question of entanglement at
the output of a beam splitter was investigated in very general terms
\cite{haung,kim}. Even more remarkable behavior has been shown
experimentally \cite{ou,kobayashi,nagata} and theoretically
\cite{paternostro}. For other optical elements like phase shifters a
quantum field with precise number of photons $n$ undergoes a phase
shift which is $n \phi$ whereas a classical beam undergoes a phase
shift $ \phi$. It is thus important to understand the behavior of
single photons and more generally nonclassical light at other
optical elements. Very recently coupled wave\-guide systems, which
are relatively easy to fabricate \cite{iwanow} and which are also
relatively decoherence free, have been shown to be good candidates
for continuous time random walks \cite{hagai}. The paper by Perets
et al. \cite{hagai} deals with classical beams of light. In the
light of what we have learnt with beam splitters and phase shifters
and the fact that Feynman \cite{feynman} used the term quantum walk
to describe the behavior of quantum particles, it is worth examining
how single photons and more generally nonclassical light would
behave in coupled waveguide systems. This way we would be able to
understand quantum walk  by single photons in coupled wave\-guides
\cite{aharonov, bose,jeong, pathak, do}.

In this letter we consider the system of coupled wave\-guides and
report the propagation of single photons and nonclassical light. We
discuss how nonclassical light from one wave\-guide gets transported
to other wave\-guides. This is especially important in applications
to quantum information science where one is very often interested in
the storage and retrieval of a quantum state \cite{biswas,politi}.
We further report how nonclassical light at one input port can
entangle different wave\-guides. We present analytical results for
Heisenberg operators and wave functions for fields in different
waveguides. For coupled wave\-guides, we show an analog of the
well-known Hong-Ou-Mandel two-photon interference. We also report
the amount of squeezing that can be produced in different
waveguides.

We consider an array of $N$ single-mode wave\-guides, coupled
through nearest-neighbor interaction. We will number the
wave\-guides from $1$ to $N$. The mode for the field in the $j^{th}$
waveguide is described by the annihilation (creation) operator
$a_{j}\mydag$($a^\dagger_{j}$). The operators $a_{j}\mydag$ and
$a^\dagger_{j}$ for the coupled wave\-guide system obey the boson
commutation relation $[ a_{j}\mydag,a^\dagger_{j} ]=1$.

 The Hamiltonian for the system can be written as,

\begin{eqnarray}
 H = \hbar g \sum_{j=1}^{N}   a_j^\dagger \hspace{0.05cm} a_j\mydag+  \hbar J  \sum_{j=1}^{N-1}(a_j^\dagger\hspace{0.05cm}
 a_{j+1}\mydag+ a_{j+1}^\dagger\hspace{0.05cm}
 a_{j}\mydag)~.\label{eq4}
\end{eqnarray}

\noindent In the above equation, $N$ is the number of wave\-guides
and the coupling parameter $J$ represents the rate at which the
photons are transferred to the neighboring wave\-guides. The
Hamiltonian (1) can be  diagonalized  by using the normal
co-ordinates given by

\begin{align}
a_{j}(t)& =\sum_{p=1}^{N}  b_{p} S(j,p)~,\\
 b_{p}(t) & =\sum_{j=1}^{N}  a_{j} S(j,p)~,
\end{align}

\noindent where the function $S(j,p)$ is defined as

\begin{eqnarray}
 S(j,p) \equiv \sqrt {\frac{{2}}{N+1}}\sin\Big(\frac{j p \pi}{N+1}\Big)~.\label{eq6}
\end{eqnarray}

\noindent This function satisfies the orthonormality relations

\begin{eqnarray}
 \sum_{p=1}^{N} S(n,p) S(m,p)&=& \delta_{n m},\nonumber \\
 \sum_{p=1}^{N-1} \big(S(n,p) S(m,p+1)&+& S(n,p+1) S(m,p)\big) \nonumber \\
 &=& 2 \hspace{0.05cm} \delta_{n m}\cos\Big(\frac{n  \pi}{N+1}\Big).
\end{eqnarray}

\noindent These two relations lead to the diagonalization of (1)~:

\begin{eqnarray}
 H &=& \hbar \sum_{p=1}^{N} (g +\beta_p) b_p^\dagger \hspace{0.05cm} b_p\mydag~,\nonumber\\
\beta_p &\equiv& 2 J \cos\Big(\frac{p\pi}{N+1}\Big)~. \label{eq7}
\end{eqnarray}

\noindent Using (2)-(6) we obtain the Heisenberg operators for the
field in each wave\-guide

\begin{eqnarray}
&& a_{j}(t)  =  \sum_{l=1}^{N} a_{l}(0) \hspace{0.05cm} A_{j,l}~, \nonumber\\
&& A_{j,l} \equiv \sum_{p=1}^{N} \exp[-i (g+\beta_{p})
t]S(l,p)\hspace{0.05cm}S(j,p)~.\label{eq14}
\end{eqnarray}

The coefficients $A_{j,l}$ determine all the quantum properties of
light in the coupled wave\-guide system. For example if we start
with light in the $l^{th}$ waveguide then $|A_{j,l}|^{2}$ gives the
propagation of light from the $l^{th}$ to the $j^{th}$ wave\-guide.
The mean number of photons in the $j^{th}$ waveguide would be

\begin{eqnarray}
N_{j}(t)  \equiv  N_{l}(0) |A_{j,l}(t)|^{2}~.
\end{eqnarray}

\begin{figure}[htp]
 \scalebox{0.38}{\includegraphics{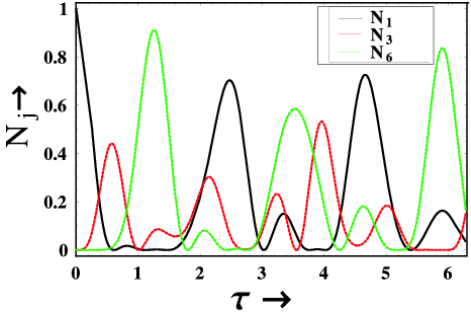}}% Here is how to import EPS art
 \caption{\label{Fig2(a)}Behavior of the normalized intensity as a function of  $\tau$ ($\tau \equiv J
\hspace{0.02cm}t/\pi $). The input is injected into the first
waveguide and the number of waveguides in the system is $N=6$}
 \end{figure}

\noindent This result is similar to the result for classical
propagation. Clearly (8) is independent of $g$. In Fig.\ 1 we show
the normalized intensity when the input is a single photon state.
Using the analytical solution (7) we can also build up the wave
function of the field at time $t$. The form of the wave function is
rather involved. For input Fock states the state at time $t$ can be
obtained in terms of the $A_{j,l}$'s given by Eq.~(\ref{eq14})~:

\begin{eqnarray}
|n_{1}, n_{2},\ldots ,n_{N}\rangle \rightarrow  \prod_{j} \frac{(
\sum A_{j,l}^*(-t) a_l^\dagger)^{n_{j}}
}{\sqrt{(n_{j})!}}|0,0,\ldots,0 \rangle~.
\end{eqnarray}

\noindent For input single photons in say waveguides $i$ and $j$ the
probability of finding one photon in the waveguide $k$ and one in
the $l^{th}$ waveguide is $|A_{i,k} A_{j,l}+A_{i,l} A_{j,k}|^{2}$~.
The two quantum amplitudes can obviously interfere. In particular
let us consider if the coupled waveguides can exhibit an analog of
Hong-Ou-Mandel interference. Consider two wave\-guides with one
photon put in each, with a delay of $T$ seconds. The wave function
at time $t$ can be related to the initial wave function using the
evolution operator $U(t)$:

\begin{eqnarray}
|\psi(t) \rangle = \frac{ U(t-T)a_{2}^\dagger U(T)
a_{1}^\dagger|0,0\rangle}{(\langle 1,0| U^\dagger(T) a_{2}\mydag
a_{2}^\dagger U(T)|1,0\rangle)^{1/2}}~.\label{eqamit}
\end{eqnarray}

\noindent In Eq.~(\ref{eqamit}) $a_{2}^\dagger$ corresponds to the
addition of a photon in the second wave\-guide at time $T$. Further
the denominator in (10) arises as we have to insure the
normalization of the wave function $a_{2}^\dagger U(T)
a_{1}^\dagger|0,0\rangle$ at time $T$. Using $U(t-T)= U(t)
U^\dagger(T)$ and the definition of the Heisenberg operators $a(t)=
U^\dagger(t) a U(t) $,  the numerator in (10) simplifies to $U(t)
a_{2}^\dagger (T) a_{1}^\dagger|0,0\rangle =  a_{2}^\dagger (T-t)
a_{1}^\dagger(-t) U(t)|0,0\rangle =  a_{2}^\dagger (T-t)
a_{1}^\dagger(-t) |0,0\rangle$.

Using the solution of Heisenberg equations in this numerator and
using (10) we find that the probability of finding simultaneously
one photon at each output at time $t$ is ($\theta = J t$,
$\theta_{o}=J T$)

\begin{eqnarray}
p (t, T) &=& |\langle 1,1|\psi\rangle|^2~, \nonumber\\
& = & \cos^2(2 \theta-\theta_{0})/(1+\sin^2(\theta_{0}))~,
\end{eqnarray}

\noindent which shows the two photon interference dip at
$2t-T=\pi/2J$ depending on the length (proportional to $t$) of the
wave\-guides and the delay time. For a given structure such a dip
can be scanned by varying the delay time.

For initial excitation in a single wave\-guide the number of photons
in each waveguide does not depend on the quantum characteristics of
the input field. We therefore examine next the squeezing and
entanglement aspects of the radiation in different wave\-guides. We
investigate the propagation of nonclassical light across the coupled
wave\-guides. We assume that squeezed light is coupled into the
first wave\-guide. The input at the first wave\-guide is

%\begin{eqnarray}
%|\zeta \rangle &\equiv& \exp\big(-\frac{r \exp(i
%\phi)}{2}(a_1^\dagger)^2+\frac{r \exp(-i \phi)}{2}(a_1)^2
%\big)|0 \rangle~, \nonumber\\
%& \equiv & \frac{1}{\sqrt{\cosh r}} \sum_{n=0}^{\infty}
%\frac{\sqrt{(2 n)!}}{2^{n} n!} {(-\exp(i \phi)\tanh
%(r))}^n |2 n\rangle~.\nonumber\\
%\end{eqnarray}

\begin{eqnarray}
|\zeta \rangle &\equiv& \exp \left( -\frac{r }{2}( e^{i \phi}
(a_1^\dagger)^2+ e^{-i \phi}(a_1\mydag)^2)
\right)|0 \rangle~, \nonumber\\
& \equiv & \frac{1}{\sqrt{\cosh r}} \sum_{n=0}^{\infty}
\frac{\sqrt{(2 n)!}}{2^{n} n!} {\big(-\exp(i \phi)\tanh
(r)\big)}^n |2 n\rangle~.\nonumber\\
\end{eqnarray}

\begin{figure}
 \begin{tabular}{cc}
 \scalebox{0.35}{\includegraphics{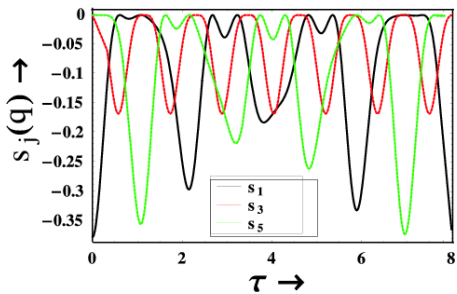}}
 \end{tabular}
 \caption{\label{Fig3}Time evolution of the squeezing factor $s_{j}(q)$ as a
function of  $\tau$ ($\tau\equiv J \hspace{0.02cm}t/\pi $). The
input is in the first waveguide and the number of waveguides in the
system is $N=5$. The magnitude and the phase of the squeezing
parameter are chosen as $r=0.7$ and $\phi=0$ respectively.}
 \end{figure}

\noindent where $r$ is the magnitude of squeezing and $\phi$ is
related to the orientation of the squeezing ellipse. In what follows
we set $g=0$. Its effect can always be incorporated by carrying out
a simple rotation.

In order to study the transport  of inter-wave\-guide squeezing, we
introduce the quadrature operators for the $j^{th}$ waveguide given
by $q_{j} \equiv (a_{j}\mydag+a^\dagger_{j})/\sqrt{2}$ and $ p_{j} \equiv
(a_{j}\mydag-a^\dagger_{j})/\sqrt{2} \hspace{0.05 cm} i$. We also define
the squeezing factors $s_{j}(q)\equiv (\Delta q_{j})^{2}- 1/2$ and
${}s_{j}(p) \equiv (\Delta p_{j})^{2}- 1/2$~. Thus squeezing occurs
when one of these expressions becomes less than zero. Using
Eq.~(\ref{eq14}) we get

\begin{eqnarray}
s_{j} = |A_{j,l}|^2 \sinh^2 r \mp \frac 1 4 \sinh{2 r}\big(A_{j,l}^2
\exp(i \phi)  + \mbox{c.c.}\big) .
\end{eqnarray}

\noindent where $-$$(+)$ sign is to be used for the quadrature q(p).

\noindent In particular for a system of two wave\-guides, we have

\begin{eqnarray}
s_{1}(q) &= &  \cos^2(J
t)\sinh(r)\big(\sinh(r)-\cos(\phi)\cosh(r)\big)~,\nonumber\\&= &
s_{2}(p) \cot^2(J t)
\end{eqnarray}

\noindent Clearly $q$-quadrature is initially squeezed if $\tanh(r)
<\cos(\phi)$. Note that for two coupled wave\-guides we obtain
complete transfer of squeezing albeit from $q$-quadrature to
$p$-quadrature for $J t=\pi/2$ \cite{footnote1}.

\noindent For a system of three wave\-guides, we get the following
results
\begin{eqnarray}
&& s_{1}(q) =  f \cos^4\Big(\frac{J t}{\sqrt{2}}\Big)~, \nonumber\\
&& s_{3}(q)  =  f \sin^4\Big(\frac{J t}{\sqrt{2}}\Big)~, \nonumber\\
&& s_{2}(p)
=  \frac{f}{2}\sin^2(\sqrt{2}J t)~,\nonumber\\
&& f  \equiv  \sinh(r)\big(\sinh(r)-\cos(\phi)\cosh(r)\big)
\end{eqnarray}

\begin{figure}
 \begin{tabular}{cc}
 \scalebox{0.4}{\includegraphics{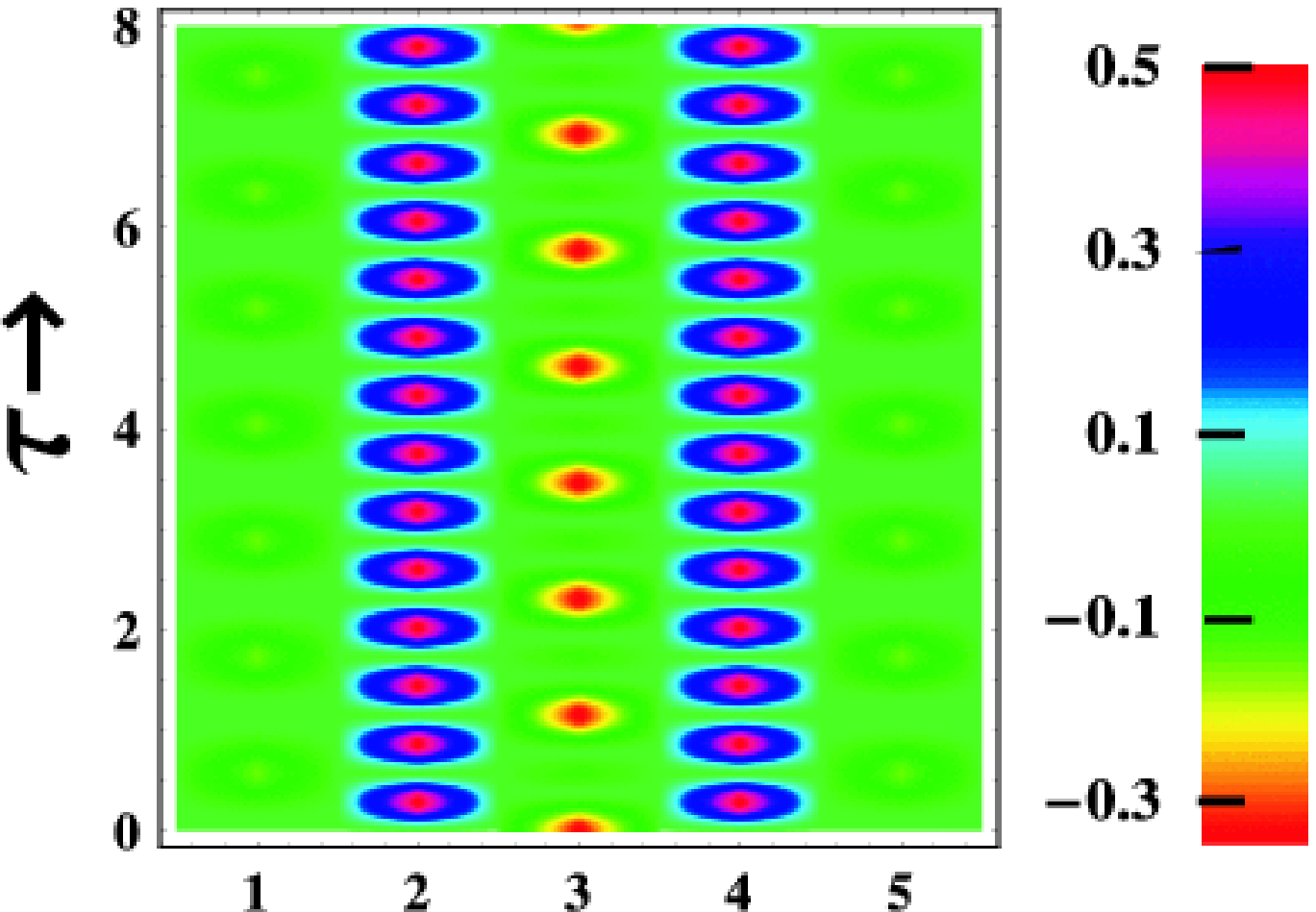}}\\\\
 \scalebox{0.4}{\includegraphics{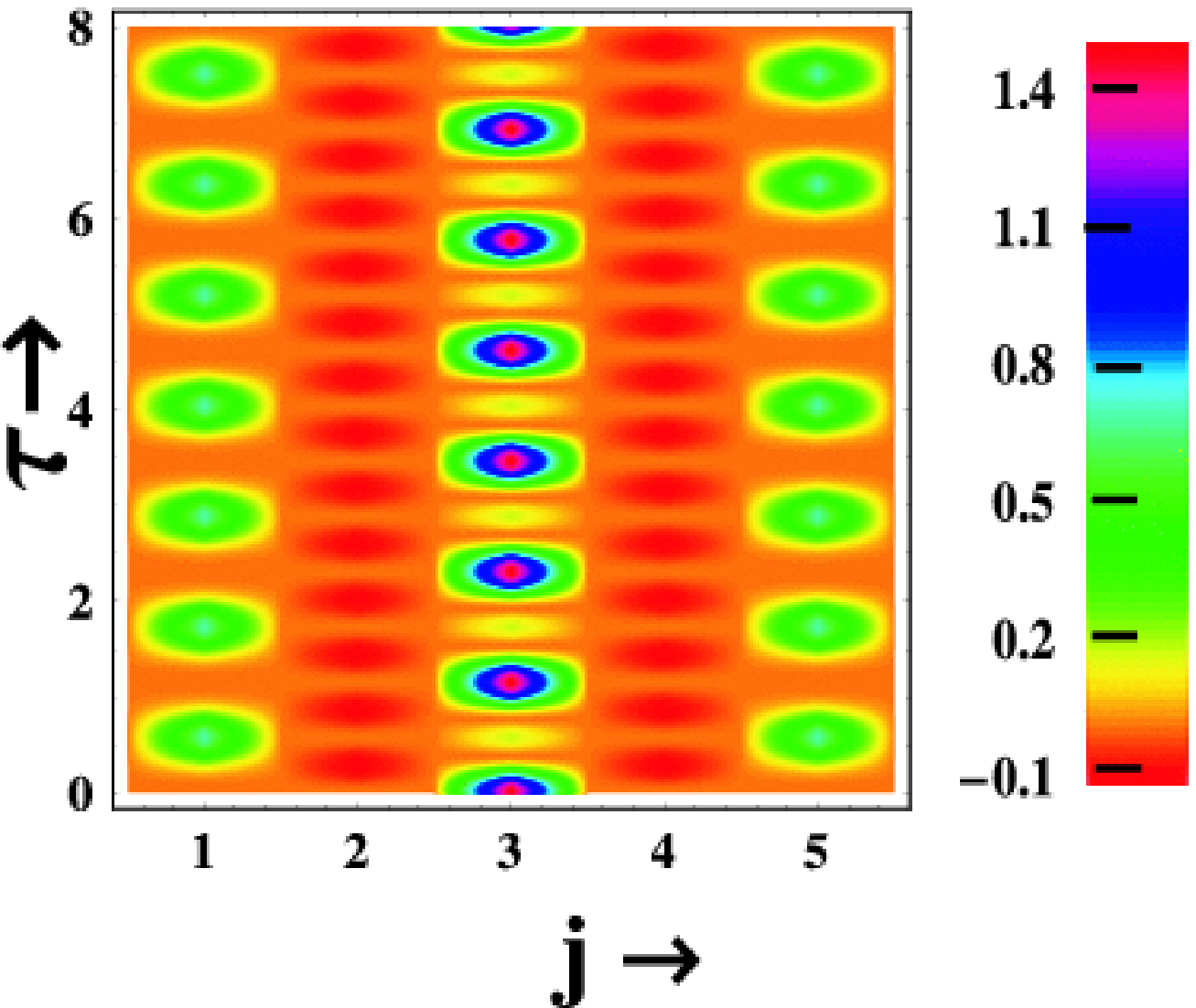}}
 \end{tabular}
 \caption{\label{Fig3}The top (bottom) part shows the variation of \-$s_{j}(q)$ ($s_{j}(p)$) as a
function of  $\tau \equiv J \hspace{0.02cm}t/\pi $ for $j=1,\ldots,5$,
smeared out in the $j$ direction. The magnitude and the phase
of the squeezing parameter are chosen as $r=0.7$ and $\phi=0$
respectively. The number of waveguides in the system is $N=5$.}
 \end{figure}

\noindent The inter-wave\-guide transfer of squeezing is governed by
the factors $\cos^4(J t/\sqrt{2})$, $\sin^4(J t/\sqrt{2})$, and $2
\sin^2(J t/\sqrt{2})\cos^2(J t/\sqrt{2})$. Again, the $q$-quadrature is
initially squeezed if $\tanh(r) <\cos(\phi)$. Also, for the case of
three coupled wave\-guides we obtain complete transfer of squeezing
from the first wave\-guide to the third wave\-guide for $J
t=\pi/\sqrt{2}$. In Fig.\ 2, we display the time evolution of the
quadrature squeezing for the case of five wave\-guides. The negative
values of $s_{q}(1)$, $s_{q}(3)$, and $s_{q}(5)$ clearly demonstrate
the squeezing in the $q$-quadrature. Fig.\ 3 shows the quadrature
squeezing when the input is given to the middle waveguide.

We next show that input of nonclassical light to one of the
wave\-guides can produce pairwise entanglement between the
inter-waveguide modes. We use the well-known criterion for
entanglement between two continuous variable systems
\cite{duan,simon,mancini}. As a measure of entanglement we examine
the correlation between two waveguide modes.
\begin{eqnarray}
M(j,k) = \langle a_j^\dagger a_{j}\mydag\rangle+ \langle a_k^\dagger
a_{k}\mydag\rangle+\langle a_j\mydag a_{k}\mydag\rangle+\langle a_j^\dagger
a_k^\dagger\rangle.
\end{eqnarray}

\noindent The negativity of $M$ is a sufficient condition for
entanglement. For Gaussian states this is both necessary and
sufficient. A calculation shows that the joint state of the coupled
wave\-guides is Gaussian. We calculate $M$ using the solution (8)
and the state in (12). Before we show numerical results, we discuss
analytical results for two and three wave\-guides. In particular,
for a system of two wave\-guides,
\begin{equation}
M(1,2) = \frac{1}{2}(\sinh(2 r)(\tanh r-\sin(2Jt)\sin \phi))
\end{equation}
and thus entanglement occurs for $\sin(2J
t)\sin(\phi)>\tanh(r)$.

For a system of three wave\-guides, we find the results

\begin{eqnarray}
M(1,2) & =& \frac{1}{2}\cos^2 \Big(\frac{J t}{\sqrt{2}}\Big)
\Big(\big(3-\cos(\sqrt{2} J t)\big)\sinh^2(r) \nonumber \\ && - \sqrt{2}
\sin(\phi)\sinh(2r)\sin(\sqrt{2} J
t) \Big)~,\\
M(1,3) & = & \frac{1}{4} \Big( \cos(\phi)\sinh(2r)\sin^2(\sqrt{2}
J t)  \nonumber  \\
&&  + \big(3+\cos(2 \sqrt{2}J t)\big)\sinh^2(r) \Big).
\end{eqnarray}

\begin{figure}
 \begin{tabular}{cc}
\scalebox{0.35}{\includegraphics{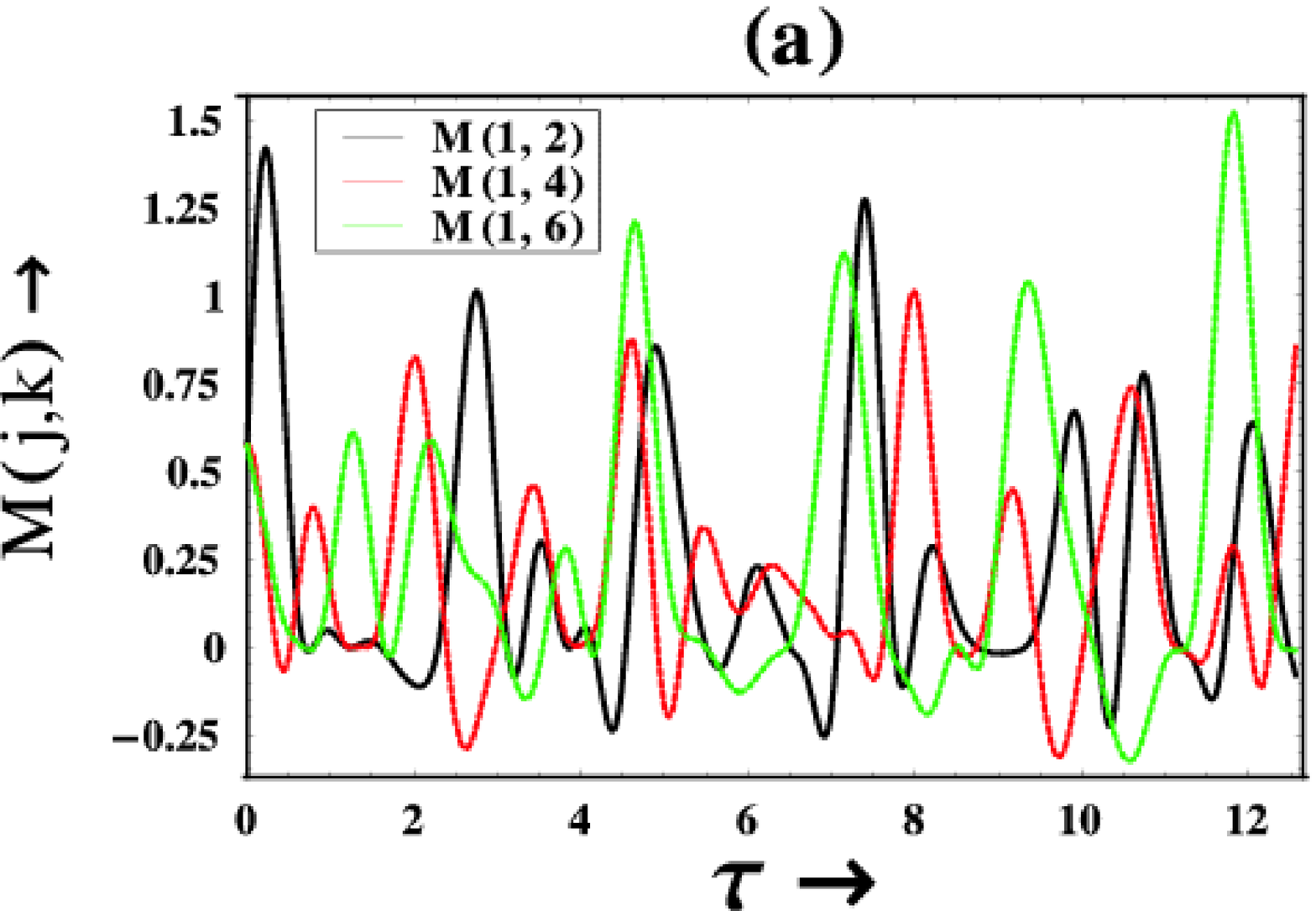}}\\
\scalebox{0.35}{\includegraphics{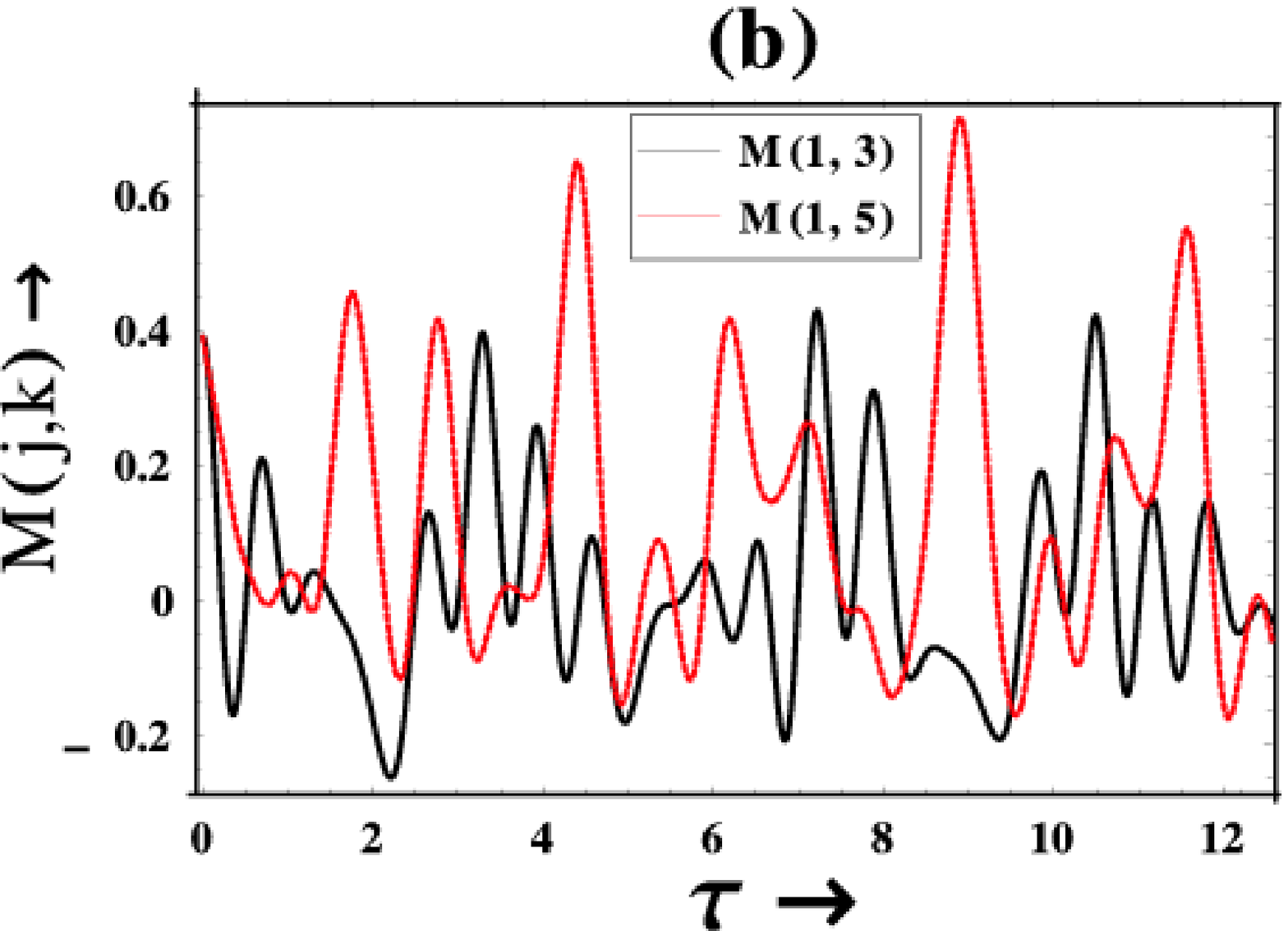}}
 \end{tabular}
 \caption{\label{Fig3}The correlation function $M(j,k)$ as a function of $\tau$ ($\tau \equiv J \hspace{0.02cm}t/\pi $) for the case of six waveguides. The
magnitude and the phase of the squeezing parameter are chosen as (a)
$r=0.7$ and $\phi=3 \pi/2$ ; (b) $r=0.6$ and $\phi=\pi$
respectively. }
 \end{figure}

\noindent Clearly for $\phi=\pi$, the first and third could be
entangled. In Fig.\ 4, we show the time evolution of  $M(j,k)$ for
the case of six wave\-guides. The negative values of $M(1,j)$ ($j=1$
to $6$) in Figs.\ 4(a) and 4(b) clearly demonstrate the entanglement
between the inter-wave\-guide modes.

\indent  In conclusion, we discussed the continuous time quantum
walk of quantum particles (photons) in a physical system consisting
of $N$ coupled wave\-guides. We showed that the quantum walk with a
light source having strong quantum character can produce
entanglement between different wave\-guides in the array. We also
studied the transport of nonclassical light across the coupled
wave\-guides. We could investigate several other interesting
nonclassical situations, for example effect of launching a
distribution of entangled photon pairs into the array and we would
then have an analogy of the Talbot effect in second quantized
set-up.

\end{document}